\documentclass[review,sort&compress,number]{elsarticle}

\usepackage[T1]{fontenc}
\usepackage[utf8]{inputenc}
\usepackage{lmodern}
\usepackage{microtype}
\usepackage[english]{babel}
\usepackage{enumitem}
\usepackage{amsmath}
\usepackage{amssymb}
\usepackage{graphicx}
\usepackage{wrapfig}
\usepackage{float}
\usepackage{xcolor}
\usepackage{algorithm}
\usepackage{algorithmic}
\usepackage{multirow}
\usepackage{bbm}
\usepackage{gensymb}
\usepackage{textcomp}
\usepackage{gensymb}
\usepackage{lineno}
\usepackage{comment}

\usepackage{booktabs}

\usepackage{fancyhdr}
\begin{document}

\begin{frontmatter}
\title{Wavelet variance scale-dependence as a dynamics discriminating tool  in high-frequency urban wind speed time series}
\author{Fabian Guignard$^{1}$, Dasaraden Mauree$^2$, Mikhail Kanevski$^1$, Luciano Telesca$^3$ }
\address{$^1$IDYST, Faculty of Geosciences and Environment, University of Lausanne, Switzerland. \\
$^2$Solar Energy and Building Physics Laboratory, Ecole Polytechnique Fédérale de Lausanne, Switzerland. \\
$^3$CNR, Istituto di Metodologie per l’Analisi Ambientale, Tito (PZ), Italy. \\
Corresponding author: Fabian.Guignard@unil.ch}

\begin{abstract}

High frequency wind time series measured at different heights from the ground (from 1.5 to 25.5 meters) in an urban area were investigated by using the variance of the coefficients of their wavelet transform. 
Two ranges of scales were identified, sensitive to two different dynamical behavior of the wind speed: the lower anemometers show higher wavelet variance at smaller scales, while the higher ones are characterized by higher wavelet variance at larger scales.  
Due to the relationship between wavelet scale and frequency, the results suggest the existence of two frequency ranges, where the wind speed variability change according to the position of the anemometer from the ground.

This study contributes to better understanding of the high frequency wind speed in urban areas and to a better knowledge of the underlying mechanism governing the wind fluctuations at different heights from the ground in particular in urban area.

\end{abstract}


\end{frontmatter}

\section{Introduction}

Urban built-up and layout are strongly influenced by wind (and vice-versa), which is a crucial factor to consider in urban planning. 
For instance, an important hazard for pedestrians is represented by wind over-speed and vortices occurring in connecting passage ways between two buildings \cite{Dutt1991}. 
High quality and high frequency wind data that are registered through experimental campaigns and field trips are necessary in understanding the impact of urban areas and/or buildings on wind and for representing land surface in the evaluation of building energy use \cite{Mauree2017a, Mauree2018}, dispersion of air pollutants, renewable energy potential in urban planning scenarios \cite{Perera2018}. 
Meteo-climatic stations, in general, record also wind, but not with sufficiently high vertical resolution and high frequency. 
Campaigns such as  BUBBLE \cite{Rotach2005} provided useful data for development  and generalization of new parameterization schemes. 
However, more data of such type over longer time periods and in different  urban settings are needed, in order to develop  new methods  to evaluate  building energy use. 
For instance, the vertical profiles of wind speed close to buildings are necessary to determine the momentum and heat fluxes \cite{Mauree2017b, Jarvi2018, Santiago2007}. 
Moreover, only high frequency wind speed data could allow the identification of small turbulence structures, which often characterize urban configurations \cite{Christen2009}.

The extremely complex interactions between mean wind speed vertical gradient, turbulence, shape, size, layout of buildings, etc.\ cause  high variability at any timescales in wind speed. Therefore, robust time series methodologies are necessary to understand the temporal variations of wind speed in urban areas at different heights from the ground \cite{Mauree2017d, Mauree2017c}. 
It has previously been demonstrated either with measurements in real cases \cite{Mauree2017d, Rotach2005} or in wind tunnels \cite{Allegrini2013} or with computational methods \cite{Santiago2007} that there is a significant difference in the dynamics of the wind flow in the urban canyon and above it. 
It would thus be particularly useful to develop new tools for analyzing datasets collected in urban areas to derive new methods or models that would better represent the underlying physical processes. 

\begin{figure}
\centering
\includegraphics[width=\linewidth]{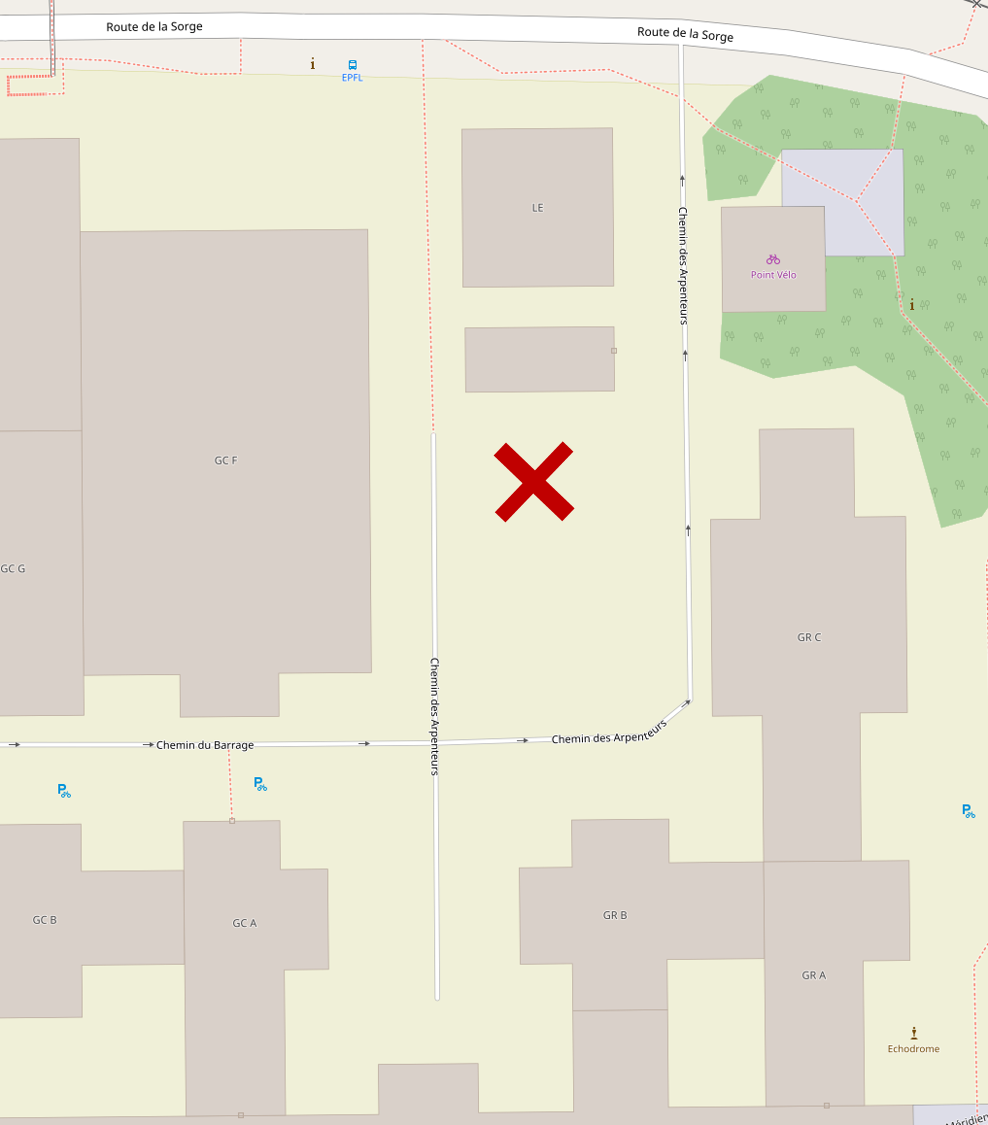}
\caption{Location of the mast.}
\label{Mast}  
\end{figure}

In the present work, seven high-frequency urban wind speed series are analyzed.
They are measured at different heights from the ground (from 1.5 to 25.5 meters, with 4 meters spacing among the sensors) on a 27 meters high mast installed on the campus of Ecole Polytechnique Fédérale de Lausanne (EPFL), Switzerland (motus.epfl.ch), whose average building height is around 10 meters \cite{Mauree2017d, Mauree2017c} (see Fig. \ref{Mast}).
The experiment was carried out to quantify the influence of urban buildings on wind fluctuations. 
Since building layouts could be considered as the main source of local turbulence phenomena in the wind flow below the average building height, the  experiment aimed at distinguishing the temporal dynamics of wind speed recorded at levels below the average building height from that of wind speed recorded at levels above. 
As a discriminating tool, wavelet variance is applied. This method allows the decomposition of the variance of each signal by scale through multiresolution wavelet analysis, which proved to be a performing tool for extracting scale-dependent components of a signal \cite{Addison2002}.

The paper is organized as follows. 
First, the technical details of the experiment are briefly reported. 
Then the method of the wavelet variance is described.
Next, the results are presented, and the final remarks are summarized in the conclusions.

\section{Description of the experiment}

\begin{figure}
\centering
\includegraphics[width=\linewidth]{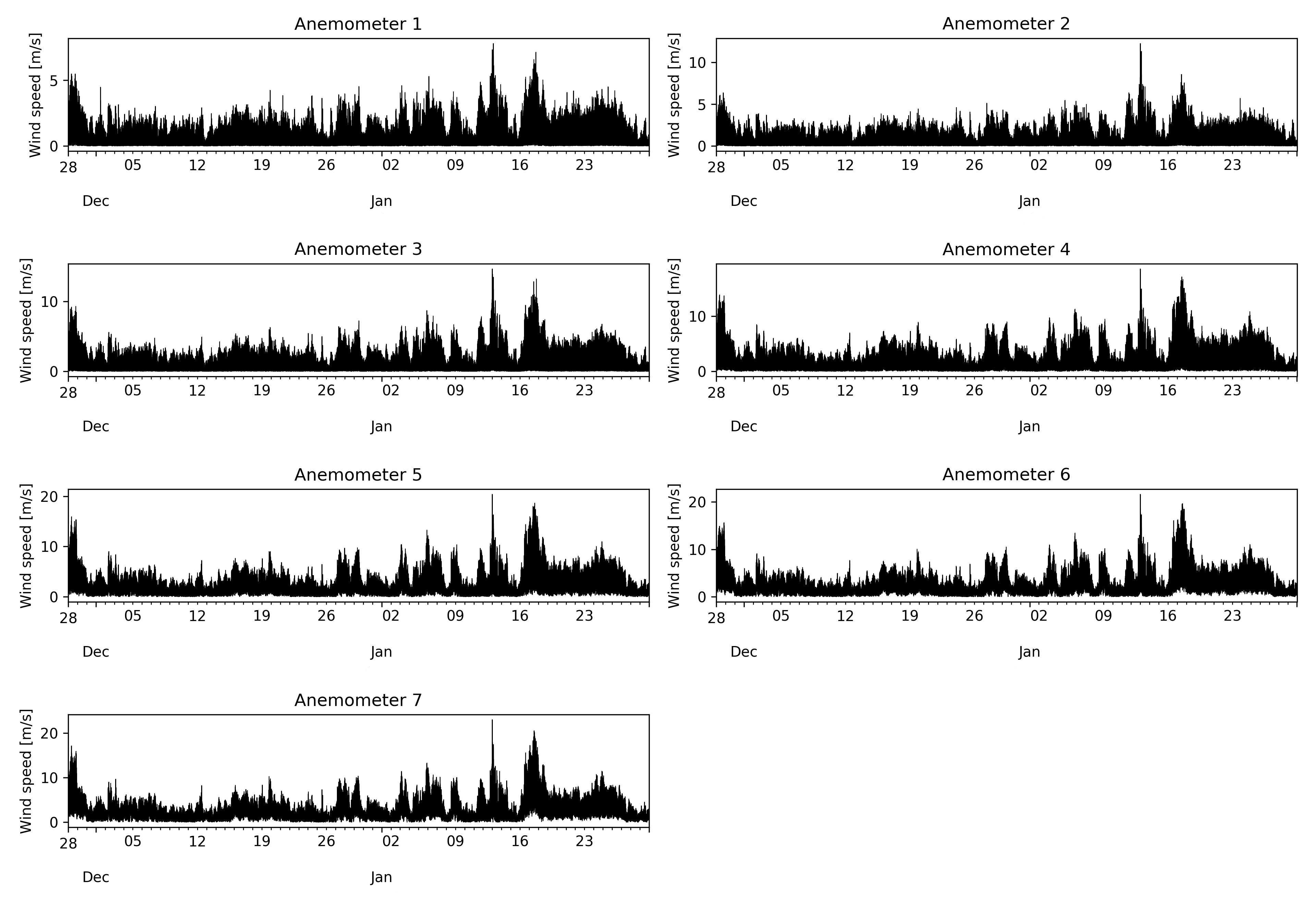}
\caption{$1 Hz$ wind speed time series for the 7 anemometers.}
\label{Data}  
\end{figure}

\begin{figure}
\centering
\includegraphics[width=\linewidth]{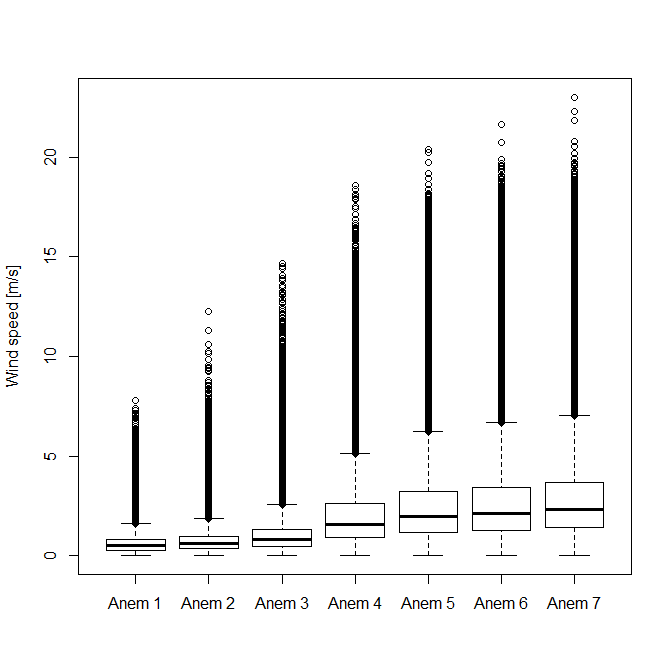}
\caption{Boxplots  for the 7 anemometers.}
\label{Boxplot}  
\end{figure}

Seven 3D sonic anemometers have  been installed along the vertical axis of a 27 m high mast every 4 m.  
The anemometers were set at 1.5 m, 5.5 m, 9.5 m, 13.5 m., 17.5 m, 21.5 m and 25.5 m above the ground. The last sensor was placed high enough above the building layout height in order to be in a undisturbed flow \cite{Rotach1999}. 
The frequency of the data is $1Hz$ \cite{Kaimal1994}. 
Each sensor registers data for the three velocity components, the sonic speed and temperature.  
The data, collected durring two months period from 28 December 2016 to 29 January 2017, are shown in Fig. \ref{Data}.
Boxplots  show a certain amount of extreme values, typical of wind data, see Fig. \ref{Boxplot}.

\section{The wavelet variance}

\subsection{Multiresolution wavelet analysis}
An accurate signal description in time and in frequency cannot be performed in the Fourier analysis framework.
The wavelet transform overcomes this problem by decomposing the signal on an orthogonal basis, which is generated by translating and scaling a function $\psi(t)$ called \textit{mother wavelet}.

\begin{figure}
\centering
\includegraphics[width=\linewidth]{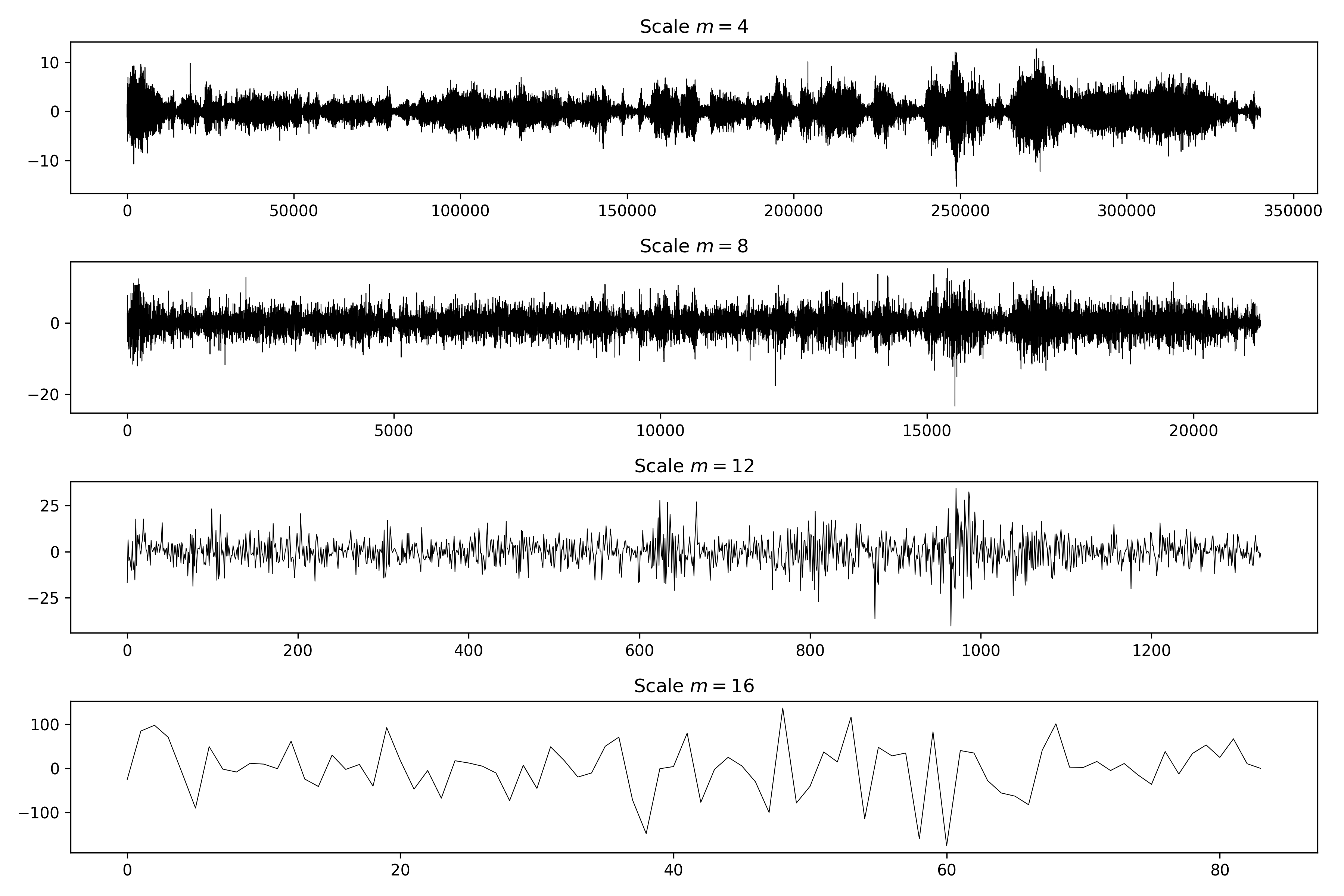}
\caption{Detailed coefficients of the DWT at scales $m = 4, 8, 12$ and $16$ for the anemometer $1$, using Haar mother wavelet.}
\label{coef2D}  
\end{figure}

\begin{figure}
\centering
\includegraphics[width=\linewidth]{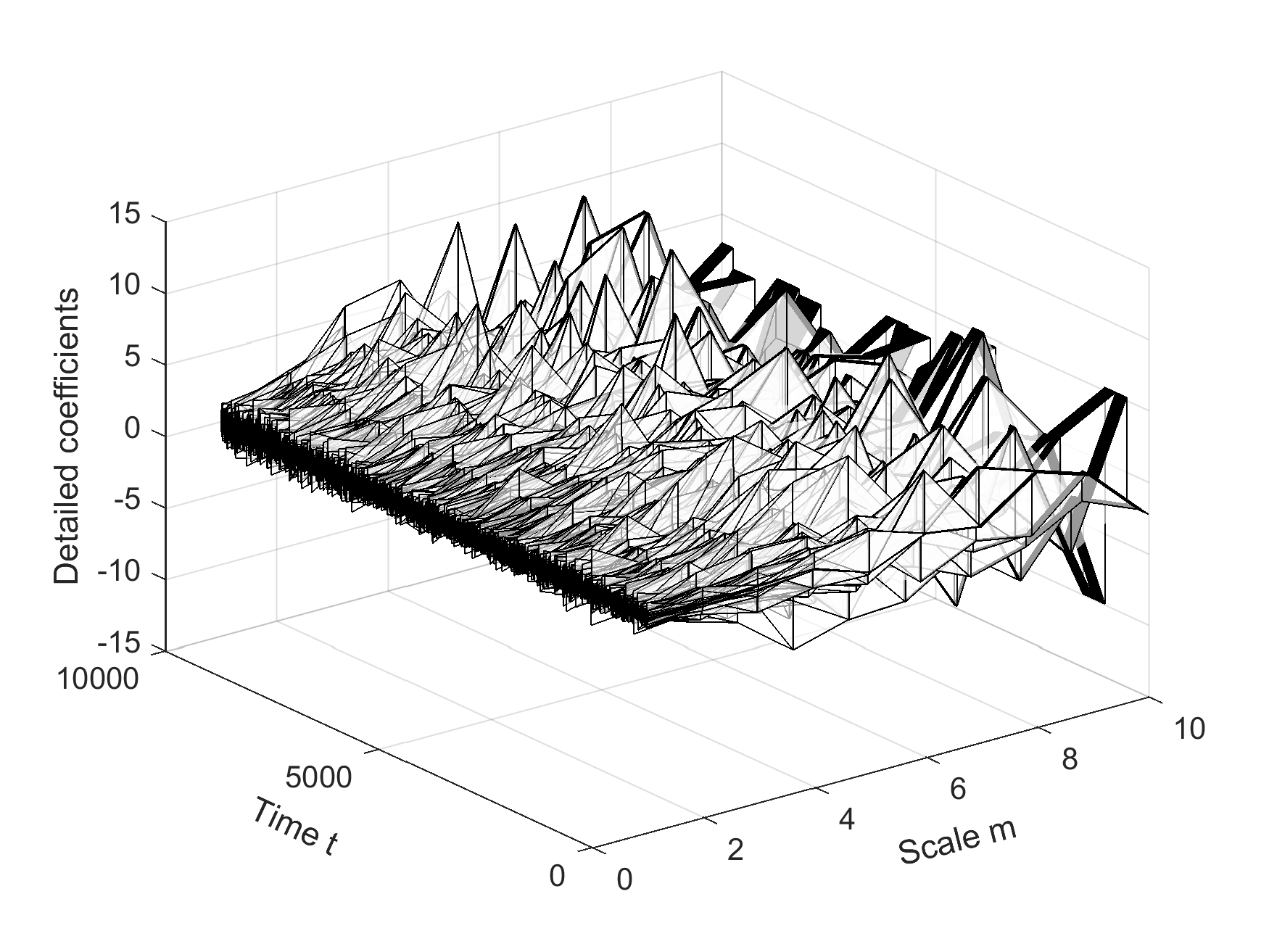}
\caption{3D representation of the DWT detailed coefficients as a function of scale over a part of the data, for the anemometer $1$, using Haar mother wavelet.}
\label{coef3D}  
\end{figure}

Multiresolution wavelet analysis (MRWA) is a recursive method for performing discrete wavelet analysis \cite{Gao2007}. 
Given a signal $x = \{x(i)\}$ of length $L$, sampled at regular intervals $\Delta t = \tau$, one can split it in two components; 1) an approximated part $A_1$ of $x$ at the coarser scale $\Delta t =2\tau$ and 2) a detailed part $D_1$ at scale $\Delta t = \tau$. This procedure is recursively performed on the approximated parts, each step yielding an approximation $A_m$ at  scale $ \Delta t =2^m \tau$ and a  detailed part $D_m$ at  scale $\Delta t = 2^{m-1} \tau$. After $M$ iterations of the process, the signal is decomposed as
\begin{equation}
x = A_M + D_1 + D_2 + \dots + D_M.
\end{equation}
In practice, this procedure is computed by performing the Discrete Wavelet Transform (DWT) algorithm \cite{Daubechies1992}.
Each detailed level $D_m$ have $N_m = int(L/2^m)$ coefficients  completely determined by the mother wavelet $\psi(t)$ and given by
\begin{equation}
d(m, n) = 2^{-m/2}\sum_{i=0}^{L-1} x(i)\psi(2^{-m} i - n),
\end{equation}
for all $m= 1, \dots, M$, $n= 0, \dots, N_m-1$ \cite{Thurner1997}.
An example of the detailed coefficients at different scale levels for Haar mother wavelet is shown in Fig. \ref{coef2D} and Fig. \ref{coef3D}.

\subsection{Wavelet variance}

In the present work, the quantity of interest is the variance of the detailed coefficients for a given scale $m$, noted as $\sigma_{det}^2(m)$. 
An estimate for the variance of the detailed coefficients is defined by
\begin{equation}\label{sigma_wav}
\hat \sigma_{det}^2(m) = \frac{1}{N_m} \sum_{n=0}^{N_m-1} \left[ d(m,n) -  \frac{1}{N_m} \sum_{n=0}^{N_m-1} d(m,n)   \right]^2,
\end{equation}
for a given scale $m$.

However, it is convenient to normalize $\hat \sigma_{det}(m)$ by $2^m$, yielding a scale decomposition of the signal variance.
Following \cite{Lindsay1996}, let's define the \textit{wavelet variance} as $\hat \sigma^2_{wav}(m) ={\hat \sigma^2_{det}(m)}/{2^m}$.
Suppose for convenience that $L =2^M$.  If the coefficient averages are zeros, then we have for all scale $m$,
\begin{align}\label{sigma_dec}
\hat \sigma^2_{wav}(m) &= \frac{\hat \sigma^2_{det}(m)}{2^m} = \frac{N_m}{L}  \hat \sigma^2_{det}(m) = \frac{1}{L} \sum_{n=0}^{N_m-1} d(m,n) ^2,
\end{align}
since $N_m = L/2^m$ and using  Eq. (\ref{sigma_wav}). Assuming the orthonormality of the basis induced by the mother wavelet, we can show \cite{Lindsay1996, Addison2002} that
\begin{equation}\label{power}
\sum_{i=0}^{L-1}x(i)^2 = \sum_{m=1}^M\sum_{n=0}^{N_m-1} d(m,n)^2 + L \bar x^2.
\end{equation}
Then, using Eq. (\ref{sigma_dec}) and (\ref{power}), we have 
\begin{align}
\sum_{m=1}^M \hat \sigma^2_{wav}(m) &= \frac{1}{L} \sum_{m=1}^M  \sum_{n=0}^{N_m-1} d(m,n)^2 =\frac{1}{ L} \sum_{i=0}^{L-1}x(i)^2 -  \bar x^2. 
\end{align}
The right-hand side member of that equation is nothing else than the sample variance of the signal. Thus, the detailed coefficients $d(m, n)$ of MRWA provide a natural scale decomposition of the variance by $\hat \sigma^2_{wav}(m)$.

\section{Results and discussion}

In this study, seven time series of $1Hz$ wind speed  measured from 1.5 m to 25.5 m from the ground at a 27.5 m mast located in the campus of EPFL were analysed, see Fig. \ref{Data}.  
The wind speed measured by the three lowest anemometers would be more affected by the canopying and tunneling effect produced by the layout of the buildings around the mast, while the higher four would be in conditions of undisturbed flow. 
These results are in agreement with previous studies such as \cite{Santiago2007, Mauree2017b, Christen2009}.
It can be clearly seen that all the seven time series seem to  be characterized by similar variability, although the amplitude of the variation increases with the height from the ground. 
The close similarity among the wind series is confirmed by the Pearson correlation analysis (Table \ref{Pearson}), which shows a correlation coefficient ranging between 0.55 and 0.96, indicating a high shape similarity.
Although such very close shape similarity among the series is observed, the wind speed series are characterized by different local forcings governing the  variability. 
In fact, the wind speed variability up to 9.5 m (a bit below the  average height of the buildings around the mast) should be more informative on the small turbulence dynamics taking place around building, which is typical of the roughness sublayer \cite{Fisher2005, Oke1976}.
The wind speed variability from 17.5 m up to 25.5 m (so, sufficiently above the average height of the building layout) would be in a status of unconditioned flow, which is typical of the inertial layer \cite{Fisher2005, Oke1976}.  
The wind speed measured at 13.5 m could be probably in an intermediate status, and thus could be affected by both phenomena. In particular at this height, the rooftop could influence the turbulent structure creating more disturbed flow above the building height \cite{Coceal2007}. 
Additionally, wake turbulence could also be responsible for increased variance at this height \cite{Oke1988}.

\begin{table}
\centering
\begin{tabular}{lrrrrrrr}
\toprule
 &     An 1 &     An 2 &     An 3 &     An 4 &     An 5 &     An 6 &     An 7 \\
\midrule
An 1 &  1.000 &  0.641 &  0.550 &  0.600 &  0.628 &  0.634 &  0.637 \\
An 2 &  0.641 &  1.000 &  0.620 &  0.574 &  0.590 &  0.597 &  0.601 \\
An 3 &  0.550 &  0.620 &  1.000 &  0.677 &  0.625 &  0.616 &  0.612 \\
An 4 &  0.600 &  0.574 &  0.677 &  1.000 &  0.895 &  0.854 &  0.830 \\
An 5 &  0.628 &  0.590 &  0.625 &  0.895 &  1.000 &  0.947 &  0.915 \\
An 6 &  0.634 &  0.597 &  0.616 &  0.854 &  0.947 &  1.000 &  0.959 \\
An 7 &  0.637 &  0.601 &  0.612 &  0.830 &  0.915 &  0.959 &  1.000 \\
\bottomrule
\end{tabular}
\caption{Pearson correlation between anemometers.}
\label{Pearson} 
\end{table}

\begin{figure}
\centering
\includegraphics[width=\linewidth]{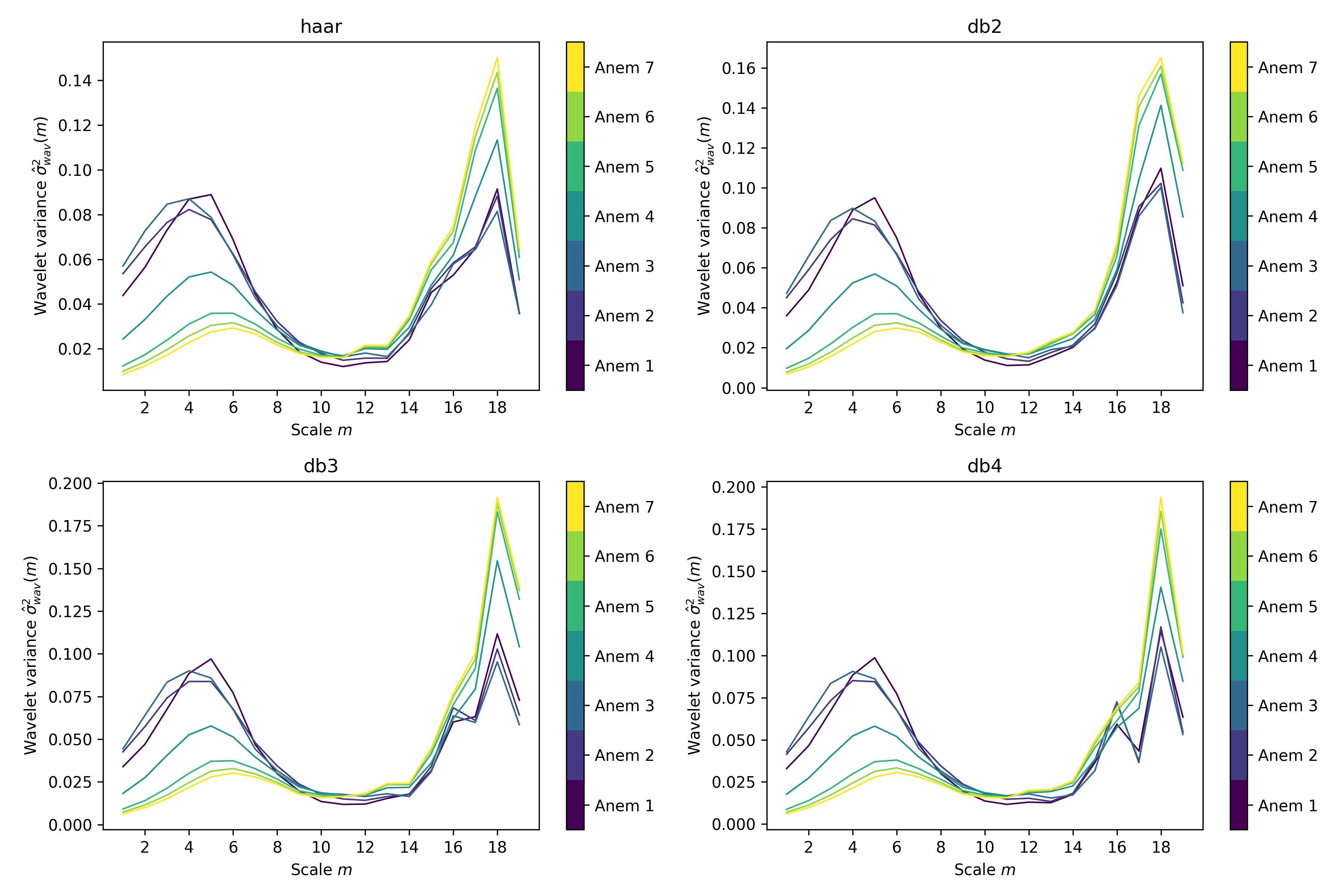}
\caption{Wavelet variance for different mother wavelets performed on the normalized time series.}
\label{Results}  
\end{figure}

The wavelet variance was applied to all seven normalized time series. 
Four different mother wavelets were used:  Haar, db2, db3 and db4.
The wavelet variance for the seven wind speed series is shown in Fig. \ref{Results}.  
All wavelet variance plots present very similar characteristics:
\begin{enumerate}
\item The wavelet variance curves versus scale $m$ of the wind speed measured by the first three lower anemometers are almost overlapped, showing very slight difference between each other;
\item The wavelet variance curves versus scale $m$ of the wind speed measured by the last three higher anemometers, similarly to the first three lower ones, are also overlapping, with very small differences between them;
\item The wavelet variance of the wind speed of the first three lower anemometers is larger at small scales, with a maximum between scales 4 and 5; while the wavelet variance of the wind speed of the last three higher anemometers is larger at larger scales, with a maximum at the scale 18;
\item The wavelet variance curve corresponding to the wind speed measured at 13.5 m seems almost equidistant between the two groups of the wind speed measured by the three lower and the three higher anemometer.
\end{enumerate}

The wavelet variance curves are in a good agreement with previous studies \cite{Stull1988}. It is clear that the lower anemometers captures intra-canyon dynamics while above the urban canyon, the synoptic flow is most important. This could relate also to the size of the turbulent eddies. Close to the ground the eddies are rather small \cite{Santiago2010, Mauree2017b} while above the ground they increase linearly with height.

\section{Conclusion}

In this study the high frequency wind speed time series, recorded on a 27.5 m high mast in a urban area are studied by means of the wavelet variance. Wavelet variance is a well-known time series analysis tool, that decomposes the variance of a series into a set of values depending on the scale. 
The investigated heights from the ground, from 1.5 m to 25.5 m, were set to discriminate the wind dynamics below and above the average building height around the mast, and to quantify its influence on a wind temporal variability.
Our findings indicate that the building layout influences the temporal variability of the wind : wavelet variance increases at smaller or larger scales for the lower or higher anemometers correspondingly. 
Considering the relationship between wavelet scale and frequency (small/large wavelet scales correspond to high/low frequency ranges), the lower anemometers are characterized by the dominance of high frequency fluctuations.
 This is in agreement with the typical phenomena of small turbulence dynamics, that takes place around buildings at small heights from the ground and that is mainly evident at high frequencies. 
Indeed, close to the ground and in proximity of buildings, the mechanical generation of turbulence is prevalent and the airflow movements are dominated by small turbulent eddies \cite{Coceal2004, Xie2008}. 
The wind measured by the higher anemometers, situated in a status of unconditioned flow, typical of inertial layer, is not characterized by intermittency or  bursts of high frequency fluctuations like the wind measured at lower heights; therefore its dynamics is more “regular” and dominated by low frequency fluctuations, which is more representative of the synoptic flows.

\section{Acknowledgements}

F. Guignard and M. Kanevski thank the support of the National Research Programme 75 “Big Data” (PNR75) of the Swiss National Science Foundation (SNSF). 
L. Telesca thanks the support of the "Scientific Exchanges" project n° 180296 funded by the SNSF.
The MoTUS experiment was funded by EPFL and the ENAC Faculty and has been financially supported by the Swiss Innovation Agency Innosuisse and is part of the Swiss Competence Center for Energy Research SCCER FEEB\&D.
The authors are grateful to Mohamed Laib, Federico Amato and Jean Golay for the profitable discussions.


\bibliography{xampl}

\begin{thebibliography}{10}
\expandafter\ifx\csname url\endcsname\relax
  \def\url#1{\texttt{#1}}\fi
\expandafter\ifx\csname urlprefix\endcsname\relax\def\urlprefix{URL }\fi
\expandafter\ifx\csname href\endcsname\relax
  \def\href#1#2{#2} \def\path#1{#1}\fi

\bibitem{Dutt1991}
A.~Dutt, Wind flow in an urban environment, Environmental Monitoring and
  Assessment 19 (1991) 495 – 506.
\newblock \href {http://dx.doi.org/10.1007/BF00401336.}
  {\path{doi:10.1007/BF00401336.}}

\bibitem{Mauree2017a}
D.~Mauree, S.~Coccolo, J.~Kaempf, J.-L. Scartezzini, Multi-scale modelling to
  evaluate building energy consumption at the neighbourhood scale, PLOS ONE 12
  (2017) 1--21.
\newblock \href {http://dx.doi.org/10.1371/journal.pone.0183437}
  {\path{doi:10.1371/journal.pone.0183437}}.

\bibitem{Mauree2018}
D.~Mauree, S.~Coccolo, A.~T.~D. Perera, V.~Nik, J.-L. Scartezzini, E.~Naboni, A
  new framework to evaluate urban design using urban microclimatic modeling in
  future climatic conditions, Sustainability 10~(4).
\newblock \href {http://dx.doi.org/10.3390/su10041134}
  {\path{doi:10.3390/su10041134}}.

\bibitem{Perera2018}
A.~Perera, S.~Coccolo, J.-L. Scartezzini, D.~Mauree, Quantifying the impact of
  urban climate by extending the boundaries of urban energy system modeling,
  Applied Energy 222 (2018) 847 -- 860.
\newblock \href
  {http://dx.doi.org/https://doi.org/10.1016/j.apenergy.2018.04.004}
  {\path{doi:https://doi.org/10.1016/j.apenergy.2018.04.004}}.

\bibitem{Rotach2005}
M.~W. Rotach, R.~Vogt, C.~Bernhofer, E.~Batchvarova, A.~Christen, A.~Clappier,
  B.~Feddersen, S.-E. Gryning, G.~Martucci, H.~Mayer, V.~Mitev, T.~R. Oke,
  E.~Parlow, H.~Richner, M.~Roth, Y.-A. Roulet, D.~Ruffieux, J.~A. Salmond,
  M.~Schatzmann, J.~A. Voogt,
  \href{https://doi.org/10.1007/s00704-004-0117-9}{Bubble -- an urban boundary
  layer meteorology project}, Theoretical and Applied Climatology 81~(3) (2005)
  231--261.
\newblock \href {http://dx.doi.org/10.1007/s00704-004-0117-9}
  {\path{doi:10.1007/s00704-004-0117-9}}.
\newline\urlprefix\url{https://doi.org/10.1007/s00704-004-0117-9}

\bibitem{Mauree2017b}
D.~Mauree, N.~Blond, M.~Kohler, A.~Clappier, On the coherence in the boundary
  layer: Development of a canopy interface model, Frontiers in Earth Science 4
  (2017) 109.
\newblock \href {http://dx.doi.org/10.3389/feart.2016.00109}
  {\path{doi:10.3389/feart.2016.00109}}.

\bibitem{Jarvi2018}
L.~J\"arvi, U.~Rannik, T.~V. Kokkonen, M.~Kurppa, A.~Karppinen, R.~D.
  Kouznetsov, P.~Rantala, T.~Vesala, C.~R. Wood, Uncertainty of eddy covariance
  flux measurements over an urban area based on two towers, Atmospheric
  Measurement Techniques Discussions 2018 (2018) 1--27.
\newblock \href {http://dx.doi.org/10.5194/amt-2018-89}
  {\path{doi:10.5194/amt-2018-89}}.

\bibitem{Santiago2007}
J.~L. Santiago, A.~Martilli, F.~Mart{\'i}n, Cfd simulation of airflow over a
  regular array of cubes. part i: Three-dimensional simulation of the flow and
  validation with wind-tunnel measurements, Boundary-Layer Meteorology 122~(3)
  (2007) 609--634.
\newblock \href {http://dx.doi.org/10.1007/s10546-006-9123-z}
  {\path{doi:10.1007/s10546-006-9123-z}}.

\bibitem{Christen2009}
A.~Christen, M.~W. Rotach, R.~Vogt, The budget of turbulent kinetic energy in
  the urban roughness sublayer, Boundary-Layer Meteorology 131 (2009) 193 –
  222.
\newblock \href {http://dx.doi.org/10.1007/s10546-009-9359-5}
  {\path{doi:10.1007/s10546-009-9359-5}}.

\bibitem{Mauree2017d}
D.~Mauree, D.~S.-H. Lee, E.~Naboni, S.~Coccolo, J.-L. Scartezzini, Localized
  meteorological variables influence at the early design stage, Energy Procedia
  122 (2017) 325 -- 330, cISBAT 2017 International ConferenceFuture Buildings
  \& Districts – Energy Efficiency from Nano to Urban Scale.

\bibitem{Mauree2017c}
D.~Mauree, L.~Deschamps, P.~Bequelin, P.~Loesch, J.-L. Scartezzini, Measurement
  of the impact of buildings on meteorological variables, in: Building
  Simulation Application Proceedings, Bolzano: bu press, 2017.

\bibitem{Allegrini2013}
J.~Allegrini, V.~Dorer, J.~Carmeliet, Wind tunnel measurements of buoyant flows
  in street canyons, Building and Environment 59 (2013) 315 -- 326.
\newblock \href
  {http://dx.doi.org/https://doi.org/10.1016/j.buildenv.2012.08.029}
  {\path{doi:https://doi.org/10.1016/j.buildenv.2012.08.029}}.

\bibitem{Addison2002}
P.~S. Addison, The Illustrated Wavelet Transform Handbook, Taylor \& Francis,
  2002.

\bibitem{Rotach1999}
M.~W. Rotach, On the influence of the urban roughness sublayer on turbulence
  and dispersion, Atmospheric Environment 33~(24) (1999) 4001 -- 4008.
\newblock \href
  {http://dx.doi.org/https://doi.org/10.1016/S1352-2310(99)00141-7}
  {\path{doi:https://doi.org/10.1016/S1352-2310(99)00141-7}}.

\bibitem{Kaimal1994}
J.~C. Kaimal, J.~J. Finnigan, Atmospheric Boundary Layer Flows: Their Structure
  and Measurement, Oxford University Press, 1994.

\bibitem{Gao2007}
J.~Gao, Y.~Cao, W.-w. Tung, J.~Hu, Multiscale Analysis of Complex Time Series,
  John Wiley \& Sons, 2007.

\bibitem{Daubechies1992}
I.~Daubechies, Ten Lectures on Wavelets, Society for Industrial and Applied
  Mathematics, 1992.

\bibitem{Thurner1997}
S.~Thurner, M.~Feurstein, M.~C. Teich, Multiresolution wavelet analysis of
  heartbeat intervals discriminates healthy patients from those with cardiac
  pathology, 1997.

\bibitem{Lindsay1996}
R.~W. Lindsay, D.~B. Percival, D.~A. Rothrock, The discrete wavelet transform
  and the scale analysis of the surface properties of sea ice, IEEE
  Transactions on Geoscience and Remote Sensing 34~(3) (1996) 771--787.
\newblock \href {http://dx.doi.org/10.1109/36.499782}
  {\path{doi:10.1109/36.499782}}.

\bibitem{Fisher2005}
P.~Fisher, J.~Kukkonen, M.~Piringer, R.~M. W., M.~Schatzmann, Meteorology
  applied to urban air pollution problems: concepts from cost 715, Atmospheric
  Chemistry and Physics Discussions, European Geosciences Union 5 (2005) 7903
  -- 7927.

\bibitem{Oke1976}
T.~Oke, The distinction between canopy and boundary‐layer urban heat islands,
  Atmosphere 14~(4) (1976) 268--277.
\newblock \href {http://dx.doi.org/10.1080/00046973.1976.9648422}
  {\path{doi:10.1080/00046973.1976.9648422}}.

\bibitem{Coceal2007}
O.~Coceal, A.~Dobre, T.~G. Thomas, S.~E. Belcher, Structure of turbulent flow
  over regular arrays of cubical roughness, Journal of Fluid Mechanics 589
  (2007) 375–409.
\newblock \href {http://dx.doi.org/10.1017/S002211200700794X}
  {\path{doi:10.1017/S002211200700794X}}.

\bibitem{Oke1988}
T.~Oke, Street design and urban canopy layer climate, Energy and Buildings
  11~(1) (1988) 103 -- 113.
\newblock \href
  {http://dx.doi.org/https://doi.org/10.1016/0378-7788(88)90026-6}
  {\path{doi:https://doi.org/10.1016/0378-7788(88)90026-6}}.

\bibitem{Stull1988}
R.~B. Stull, An Introduction to Boundary Layer Meteorology, Kluwer Academic
  Publishers, 1988.

\bibitem{Santiago2010}
J.~L. Santiago, A.~Martilli, A dynamic urban canopy parameterization for
  mesoscale models based on computational fluid dynamics reynolds-averaged
  navier–stokes microscale simulations, Boundary-Layer Meteorology 137 (2010)
  417--439.
\newblock \href {http://dx.doi.org/10.1007/s10546-010-9538-4}
  {\path{doi:10.1007/s10546-010-9538-4}}.

\bibitem{Coceal2004}
O.~Coceal, S.~E. Belcher, A canopy model of mean winds through urban areas,
  Quarterly Journal of the Royal Meteorological Society 130 (2004) 1349–1372.
\newblock \href {http://dx.doi.org/10.1256/qj.03.40}
  {\path{doi:10.1256/qj.03.40}}.

\bibitem{Xie2008}
Z.-T. Xie, O.~Coceal, I.~P. Castro, Large-eddy simulation of flows over random
  urban-like obstacles, Boundary-Layer Meteorology 129.
\newblock \href {http://dx.doi.org/10.1007/s10546-008-9290-1}
  {\path{doi:10.1007/s10546-008-9290-1}}.

\end{thebibliography}
\bibliographystyle{elsarticle-num}


\newpage

\end{document}